\documentclass[amsproc]{amsproc}
\newlength{\extramargin}
\makeatletter
\usepackage{amsmath, amssymb, amsfonts}
\newtheorem{thm}{Theorem}[section] 
\newtheorem{cor}[thm]{Corollary}
\newtheorem{lem}[thm]{Lemma}
\newtheorem{defn}[thm]{Definition}

\newtheorem{preremark}[thm]{Remark}
\newenvironment{remark}%
  {\begin{preremark}\upshape}{\end{preremark}}
\newtheorem{notation}[thm]{Notation}
\numberwithin{equation}{section}

\numberwithin{equation}{section}

\newcommand{\second}{\prime \prime }

\newcommand{\ten}{\otimes}
\newcommand{\del}{\bigtriangleup}

\newcommand{\inv}{^{-1}}

\DeclareMathOperator{\Hom}{Hom}

\DeclareMathOperator{\Res}{Res}

\begin{document}

\title[Super-bicharacter construction of quantum vertex algebras]{ Super-bicharacter construction of quantum vertex algebras}
\author{ Iana I. Anguelova }
\address{Centre de Recherches Mathematiques (CRM)\\
Montreal,  Quebec H3C 3J7 \\ Canada}
\email{anguelov@crm.umontreal.ca}

\dedicatory{To Mickey}

\maketitle

\begin{abstract}
     We extend the bicharacter construction of quantum vertex algebras first proposed by Borcherds to the case of super Hopf algebras. We give a bicharacter description of the charged free fermion super vertex algebra, which allows us to construct  different quantizations of it in the sense of $H_D$-quantum vertex algebras, or specializations to Etingof-Kazhdan quantum vertex algebras. We give  formulas for the analytic continuation of product of fields, the operator product expansion and the normal ordered product in terms of the super-bicharacters.
\end{abstract}

\noindent


\vspace{0.4cm}

\section{Introduction}
Vertex operators were introduced in the earliest days of string theory
and axioms for vertex algebras were developed to incorporate these
examples (see for instance \cite{Bor2}, \cite{FLM}, \cite{Kac}).
Similarly, the definition of a quantum vertex algebra should be such that it       accommodates  the existing examples of quantum vertex operators and their properties (see for instance \cite{FJ}, \cite{FR2}  and many others). 
There are several proposals for the definition of   a quantum vertex algebra.  They include  Borcherds' theory of (A, H, S)-vertex algebras, see
\cite{Borc}, the Etingof-Kazhdan theory of quantum vertex
algebras, \cite{EK} , and the Frenkel-Reshetikhin
theory of deformed chiral algebras, see \cite{FR}. (H. Li
has developed the Etingof-Kazhdan theory further, see for example
\cite{MR2220654}, \cite{MR2215259}.)
One of the major well known differences between quantum vertex algebras and the usual nonquantized  vertex algebras is that the  quantum vertex operators can no longer satisfy a locality (or ``commutativity'') axiom, and there is instead a braiding map controlling the failure of locality. 
 In the paper \cite{AB} we introduce the notion of an $H_D$-quantum vertex
algebra (where $H_D=\mathbb{C}[D]$ is the Hopf algebra of
infinitesimal translations), generalizing the Etingof-Kazhdan theory of quantum vertex  algebras  in
various ways.  In particular, the definition of an $H_D$-quantum vertex algebra
introduces, besides the braiding map, a translation map controlling the failure of translation covariance. (Most quantum vertex operators with non-rational braiding maps do not satisfy the usual translation covariance, \cite{Ang}). An $H_D$ quantum vertex algebra essentially 
specializes to an Etingof-Kazhdan quantum vertex algebra in the case when  the translation map is  identity. (When the translation  map is  identity, one can assume the rationality of the braiding map, see \cite{AB} for a precise statement). 

In \cite{AB} we use a bicharacter construction first proposed by  Borcherds in \cite{Borc} to construct a large
  class of $H_D$-quantum vertex algebras. One particular example of
  this construction yields a quantum vertex algebra that contains the
  quantum vertex operators introduced by Jing in the theory of
  Hall-Littlewood polynomials, \cite{J1}. (The Hall-Littlewood polynomials are a one-parameter deformation of the Schur polynomials, \cite{Macd}.) The resulting $H_D$-quantum vertex algebra is a deformation of the familiar lattice vertex algebra based on the lattice $L=\mathbb{Z}$ with pairing $(m,n)\mapsto mn$, i.e.,  the bosonic part of the boson-fermion correspondence.  The goal of this paper is to extend the  bicharacter construction to the category  of super vector spaces, i.e., the case when  the vector space underlying the vertex algebra has the additional structure of a  super Hopf algebra.  This will allow us to work in particular with  quantum fields defined on the fermionic Fock space, i.e.,  with deformations of the fermionic part of the boson-fermion correspondence. 

One of the  benefits of the bicharacter construction is the fact that it provides explicit formulas for the braiding  (and translation)  map(s). Without the bicharacters, in general  formulas for the braiding map are  only   given   for proper subspaces of the quantum vertex algebra (or deformed chiral algebra, as for example in \cite{FR}). This happens when a formula for the braiding map is known for the generating fields, but not for all of their descendants. One solution to this problem is when  the braiding map  is rational, which is  what the definition of a quantum vertex algebra in the sense of Etingof and Kazhdan, \cite{EK},  assumes. But many quantum vertex operators (including the Jing vertex operators) have non-rational braiding between them,  thus leaving the bicharacter description as an only alternative so far for providing a formula for the braiding map on the whole vector space of the quantum vertex algebra.

Even for nonquantized vertex algebras the bicharacter construction has another benefit--- there are explicit formulas for the operator product expansions of fields, as well as for the normal ordered products,  in terms of the algebra product on $V$. We also have a formula for  the analytic continuation of a product of  fields.

The outline of the paper is as follows. In the next section we recall the definition of an $H_D$-quantum vertex algebra. Next we proceed to describe the super-bicharacter construction, with main result Theorem \ref{thm:h_d-quantum-vertex-from-bichar}. We give the formulas for the analytic continuation of product of fields, the operator product expansion and the normal ordered product in terms of the super-bicharacters in Lemma \ref{lem:analcont2}, Corollary \ref{cor:opes} and Corollary \ref{cor:normord}. In the last section we give a  bicharacter description of the main example---the charged free fermion vertex algebra, Theorem \ref{thm:chargedfreeferm}. That in turn allows us to obtain many quantizations in the sense of $H_D$-quantum vertex algebra (or specializations to Etingof-Kazhdan quantum vertex algebras).

\vspace{0.4cm}

\section{$H_D$-quantum vertex algebras}
In this section we recall the definition of an $H_D$-quantum vertex algebra from \cite{AB}. The definition of a (classical) super vertex algebra  can be found in many sources, for instance \cite{Kac}, therefore we will not recall it here.

Let $t$ be a variable. We will use $t$ to describe quantum
deformations, the classical limit corresponding to $t\to 0$. Let
$k=\mathbb {C}[[t]]$ and let $V$ be an $H_D$-module and free
$k$-module. Denote by $V[[t]]$ the space of (in general infinite) sums
\[
v(t)=\sum_{i=0}^\infty v_i t^i,\quad v_i\in V.
\]
 In the same way will consider
spaces such as $V[[z]][z\inv][[t]]$ consisting of sums
\[
v(z,t)=\sum_{i=0}^\infty v_i(z) t^i,\quad v_i\in V[[z]][z\inv].
\]
We will also consider rational expressions in multiple variables and
their expansions. For instance for a rational function in $z_1$, $z_2$
with only possibly poles at $z_1=0$, $z_2=0$ or $z_1-z_2=0$ we can
define expansion maps
\begin{align*}
  i_{z_1;z_2} \colon \frac1{z_1-z_2}&\mapsto
  \sum_{n\ge0}z_1^{-n-1}z_2^n, &\quad \frac 1{z_1}&\mapsto \frac
  1{z_1}, &\quad \frac 1{z_2}&\mapsto \frac 1{z_2},\\
  i_{z_2;z_1} \colon \frac1{z_1-z_2}&\mapsto
  -\sum_{n\ge0}z_2^{-n-1}z_1^n,&\quad \frac 1{z_1}&\mapsto \frac
  1{z_1},&\quad  \frac 1{z_2}&\mapsto \frac 1{z_2},\\
  i_{z_2;z_1-z_2} \colon \frac1{z_1} &\mapsto
  \sum_{n\ge0}z_2^{-n-1}(z_1-z_2)^n,&\quad \frac 1{z_2}&\mapsto \frac
  1{z_2}, &\quad \frac 1{z_1-z_2}&\mapsto \frac 1{z_1-z_2}.
\end{align*}
We will write $i_{z_1,z_2;w_1}$ for $i_{z_1;w_1}i_{z_2;w_1}$, and
$i_{z_1,z_2;w_1,w_2}$ for $i_{z_1,z_2;w_1}\allowbreak
i_{z_1,z_2;w_2}$. 

If $A\in V\otimes V$ then we define for instance
$A^{23}, A^{13}\in V^{\otimes 3}$ by $A^{23}=1\otimes A$, and
$A^{13}=a^\prime \otimes 1\otimes a^{\prime\prime}$, if
$A=a^\prime\otimes a^{\prime\prime}$.
\begin{defn}\begin{bf} ($H_D$-quantum vertex algebra)\end{bf} \label{defn:h_d-quantum-vertex-alg}
  Let $V$ be a free $k=\mathbb{C}[[t]]$-module and an $H_D$-module.
  An $H_D$-quantum vertex algebra structure on $V$ consists of
  \begin{itemize}
  \item $1\in V$, the vacuum vector.
  \item a (singular) multiplication map
\[
X_{z_1,z_2}\colon V^{\otimes 2}\to V[[z_1,z_2]][z_1\inv,
(z_1-z_2)\inv][[t]].
\]
\item A braiding map $S^{(\tau)}$ and a translation map $S^{(\gamma)}$
  of the form
  \begin{align*}
    S^{(\tau)}_{z_1,z_2}&\colon V^{\otimes 2}\to V^{\otimes2}[z_1^{\pm
      1},z_2^{\pm 1},(z_1-z_2)^{- 1}][[t]],\\
    S^{(\gamma)}_{z_1,z_2}&\colon V^{\otimes 2}\to
    V^{\otimes2}[z_1^{\pm1},z_2,(z_1+\gamma)^{\pm1},(z_2+\gamma),
    (z_1-z_2)^{-1} ][[t]] .
  \end{align*}
  \end{itemize}
These objects satisfy the following axioms:
\begin{description}
\item[(Vacuum)] For $i=1,2$
  \begin{align}
    X_{z_1,z_2}(a\ten 1)&=e^{z_1D}a,&     X_{z_1,z_2}(1\ten a)&=e^{z_2D}a, \label{eq:vacuumX}\\
S_{z_1,z_2}(a\ten 1)&=a\ten 1,& S_{z_1,z_2}(1\ten a)&=1\ten a. \label{eq:vacuumS}
  \end{align}
Here and below we write generically $S$ for both $S^{(\tau)}$ and
$S^{(\gamma)}$. 

\item[($H_D$-covariance)]
\begin{align}
    X_{z_1,z_2}(a\otimes Db)&=\partial_{z_2}X_{z_1,z_2}(a\otimes b), \label{eq:H_DcovX}\\
    (1\otimes e^{\gamma
      D})i_{z_1-z_2, z_2;\gamma}S_{z_1,z_2+\gamma}&=S_{z_1,z_2}(1\otimes
    e^{\gamma D}), \label{eq:H_DcovS} \\
    e^{\gamma D}X_{z_1,z_2}S^{(\gamma)}_{z_1,z_2} &= X_{z_1+\gamma,z_2+\gamma}\label{eq:H_DcovMult}.
  \end{align}

\item[(Yang-Baxter)] 
  \begin{equation}\label{eq:YBaxiom}
S_{z_1,z_2}^{12}S_{z_1,z_3}^{13}S_{z_2,z_3}^{23}=S_{z_2,z_3}^{23}S_{z_1,z_3}^{13}S_{z_1,z_2}^{12}.
  \end{equation}
\item[(Compatibility with Multiplication)] 
  \begin{align}
    S_{z_1,z_2}(X_{w_1,w_2}\otimes 1)
    &=(X_{w_1,w_2}\otimes1)i_{z_1, z_1-z_2;w_1,w_2}
    S_{z_1+w_1,z_2}^{23}S_{z_1+w_2,z_2}^{13},\label{eq:CompatXx1}\\
    S_{z_1,z_2}(1\otimes X_{w_1,w_2})
    &=(1\otimes X_{w_1,w_2})i_{z_1-z_2, z_2;w_1,w_2}
    S_{z_1,z_2+w_1}^{12}S_{z_1,z_2+w_2}^{13}.\label{eq:Compat1xX}
  \end{align}
\item[(Group Properties)] 
  \begin{align}
    S_{z_1,z_2}^{(\tau)}\circ \tau\circ S_{z_2,z_1}^{(\tau)}\circ \tau&=1_{V^{\otimes 2}},\label{eq:GrpSStau}\\
    S^{(\gamma_1)}_{z_1,z_2}S^{(\gamma_2)}_{z_1+\gamma_1,z_2+\gamma_1}&=S^{(\gamma_1+\gamma_2)}_{z_1,z_2},
    \label{eq:GrpSgam1gam2}\\
    S^{(\gamma=0)}_{z_1,z_2}&=1_{V^{\otimes 2}}.\label{eq:SatZero}
  \end{align}
\item[(Locality)] For all $a,b\in V$ and $k\ge 0$ there is $N\ge0$
  such that for all $c\in V$
\begin{multline} 
  (z_1-z_2)^NX_{z_1,0}(1\ten
  X_{z_2,0})(a\ten b\ten c)\equiv\\
  \equiv (z_1-z_2)^N X_{z_2,0}(1\ten X_{z_1,0})\left(i_{z_2
      ;z_1}S^{(\tau)}_{z_2,z_1}(b\ten a)\ten c\right)\mod
  t^k.\label{eq:localityAx}
\end{multline}

\end{description}
\end{defn}
We have formulated the axioms of an $H_D$-quantum vertex algebra in
terms of the rational singular multiplication $X_{z_1,z_2}$.
Traditionally the axioms of a vertex algebra have been formulated in
terms the 1-variable vertex operator $Y(a, z)$, where $a\in V$.
To make contact with the usual notation and terminology in the theory of vertex
algebras we recall  some definitions.
\begin{defn}[\textbf{Field}]\label{defn:field} Let $V$ be a
  $k$-module. A \emph{field} on $V$ is an element of 
$\Hom(V,V((z))[[t]])$.
If $a(z)$ is a field, we have for all \ $b\in V$,  
$a(z)b\in V((z))[[t]]$. 
\end{defn}
\begin{defn}[\textbf{Vertex operator}]\label{def:vertexoperatorY}
  If $V$ is an $H_D$-quantum vertex algebra we can define the vertex
  operator $Y(a,z)$ associated to $a\in V$ by
\begin{equation}
  \label{eq:DefY}
  Y(a,z)b=X_{z,0}(a\otimes b),
\end{equation}
  for $b\in V$.
\end{defn}
Note that the vertex operator $a(z)=Y(a,z)$ for an $H_D$-quantum
vertex algebra is a field, for all $a\in V$.
\begin{remark}
Due to the presence of the braiding and translation maps the axioms for $H_D$-quantum vertex algebra are more symmetric when written in terms of the singular multiplication maps $X$, rather than the vertex operators $Y$. In \cite{AB} we also give an alternative set of axioms using the vertex operators $Y$.
\end{remark}
\begin{remark} When the translation map is the identity on $V\ten V$ one gets essentially a quantum vertex algebra in the sense of Etingof and Kazhdan, \cite{EK}). When in addition the braiding map is the identity on $V\ten V$ one gets a (nonquantized) vertex algebra ( \cite{AB}).
\end{remark}
\vspace{0.5cm}

\section{Bicharacter construction of super vertex algebras}
In  the paper \cite{AB} we constructed a large class of examples of $H_D$-quantum vertex algebras using the bicharacter construction first proposed by Borcherds in \cite{Borc}. To do that we had to assume that the underlying vector space $V$
is a commutative and cocommutative Hopf algebra. In this  section we will extend that construction to the case of   super Hopf algebras ($\mathbf{Z_2}$ graded Hopf algebras). We will also indicate  how imposing extra conditions on bicharacters leads to specializations to 
quantum vertex operator algebras of Etingof-Kazhdan type or (nonquantized) super vertex algebras.

We will work with the category of super vector spaces, i.e., $\mathbb  {Z}_{2}$ graded vector spaces. The flip map $\tau $  is defined by 
\begin{equation}
\label{eq:flip}
 \tau (a\ten b) =(-1)^{\tilde{a} \cdot \tilde{b}} (b\ten a)
\end{equation}
for any homogeneous elements $a, b$ in the super vector space, where $\tilde{a}$, $\tilde{b}$ denote correspondingly  the parity of $a$, $b$.
Define also the map $\tilde{I}$ by 
\begin{equation}
\label{eq:flip2}
 \tilde{I}(a\ten b) =(-1)^{\tilde{a} \cdot \tilde{b}} (a\ten b).
\end{equation}
A superbialgebra $A$ is a superalgebra, with compatible coalgebra
structure (the coproduct and counit are algebra maps). Denote the coproduct and the counit by $\del$  and $\eta$.
A Hopf superalgebra is a superbialgebra with an antipode $S$.
For a superbialgebra $V$  we will write $\del (a)=\sum \ a^{\prime}\ten a^{\second}$ for the coproduct of $a\in V$. We will also omit the
summation symbol, to unclutter the formulas.
\begin{remark} The difference from the usual Hopf algebra is in the product on
 $H\ten H$ : the product  is defined by
\begin{equation}
\label{eq:tespr}
(a\ten b)(c\ten d)= (-1)^{\tilde{b} \cdot \tilde{c} } (ac\ten bd)
\end{equation}
for any $a, b, c, d$ homogeneous elements in $H$.
One of the  consequences of this modified product is :
\begin{displaymath}
\del (a\cdot b)=\sum \ (ab)^{\prime}\ten (ab)^{\second} = \sum \
(-1)^{\tilde{a^{\second}} \cdot \tilde{b^{\prime}} } a^{\prime} b^{\prime}\ten a^{\second} b^{\second} 
\end{displaymath}
Note also that if $a$ is odd then $\eta (a)=0$ .
\end{remark}
A supercocommutative bialgebra is a superbialgebra with 
\begin{displaymath}
\tau (\del(a))=\del(a).
\end{displaymath}
\begin{notation} Henceforth we will assume that $V$ is a   Hopf supercommutative and supecocommutative superalgebra with antipode $S$. Here and below $a, b, c$ and $d$ are homogeneous elements of $V$.
\end{notation}
\begin{defn}\begin{bf}(Super-bicharacter)\end{bf}
\label{defn:bich}
Define a bicharacter on $V$ to be
a linear map $r$ from $V\ten V$ to $k (z_1, z_2)$, such that 
\begin{align}
\label{eq:iden}
r_{z_1,z_2}(1\ten a)&= \eta (a) = r_{z_1,z_2}(a\ten 1),\\
\label{eq:multleft}
r_{z_1,z_2}(ab\ten c)&=\sum \ (-1)^{\tilde{b} \tilde{c^{\prime}} } r_{z_1,z_2}(a\ten
c^{\prime})r_{z_1,z_2}(b\ten c^{\second}),\\
\label{eq:multright}
r_{z_1,z_2}(a\ten bc)&= \sum \ (-1)^{\tilde{a^{\second}} \tilde{b} } r_{z_1,z_2}(a^{\prime}\ten b)r_{z_1,z_2}(a^{\second}\ten c).
\end{align}
We say that a bicharacter $r$ is even if $r_{z_1,z_2}(a\ten b)=0$ whenever
$\tilde{a}\neq \tilde{b}$.
\end{defn}
\begin{remark} 
\label{remark:even}
 From now on we will always work with \emph{even}  bicharacters. In most cases there  are no nontrivial arbitrary bicharacters.
For instance, the definition implies that we have on one side, using the property \eqref{eq:multright}:
\begin{align}
&r_{z_1,z_2}(ab\ten cd)=\sum \ (-1)^{\tilde{b} \tilde{(cd)^{\prime}} } r_{z_1,z_2}(a\ten
(cd)^{\prime})r_{z_1,z_2}(b\ten (cd)^{\second})=\\ &=\sum \ (-1)^{\tilde{b} \tilde{c^{\prime}}+\tilde{b} \tilde{d^{\prime}} +\tilde{c^{\second}}\tilde{d^{\prime}}} r_{z_1,z_2}(a\ten c^{\prime}d^{\prime})r_{z_1,z_2}(b\ten c^{\second}d^{\second})=\\ \label{eq:310}&=\sum \ (-1)^{\tilde{b} \tilde{c^{\prime}}+\tilde{b} \tilde{d^{\prime}} +\tilde{c^{\second}}\tilde{d^{\prime}}+\tilde{a^{\second}}\tilde{c^{\prime}}+\tilde{b^{\second}}\tilde{c^{\second}}} r_{z_1,z_2}(a^{\prime}\ten c^{\prime})r_{z_1,z_2}(a^{\second}\ten d^{\prime})r_{z_1,z_2}(b^{\prime}\ten c^{\second})r_{z_1,z_2}(b^{\second}\ten d^{\second}), 
\end{align}
for any homogeneous elements $a, b, c, d\in V$.
Similarly, using the property \eqref{eq:multleft} we have 
\begin{align}
&r_{z_1,z_2}(ab\ten cd)=\sum \ (-1)^{\tilde{(ab)^{\second}}\tilde{c}  } r_{z_1,z_2}((ab)^{\prime}\ten c)r_{z_1,z_2}((ab)^{\second}\ten d)=\\ \label{eq:311} &=\sum \ (-1)^{\tilde{a^{\second}} \tilde{c}+\tilde{b}^{\second} \tilde{c} +\tilde{a^{\second}}\tilde{b^{\prime}}+\tilde{b^{\prime}}\tilde{c^{\prime}}+\tilde{b^{\second}}\tilde{d^{\prime}}} r_{z_1,z_2}(a^{\prime}\ten c^{\prime})r_{z_1,z_2}(b^{\prime}\ten c^{\second})r_{z_1,z_2}(a^{\second}\ten d^{\prime})r_{z_1,z_2}(b^{\second}\ten d^{\second}).
\end{align}
In order to have a nontrivial bicharacter, we need these two expressions to be equal. Therefore the exponents of $(-1)$ in front of every corresponding \emph{non-zero} summand should be the same. One uses the property of  $\mathbf{Z_2}$ graded Hopf algebras, namely $\tilde{a}=\tilde{a^{\second}} +\tilde{a^{\second}}$, for any $a\in V$, thus
\begin{align*}
&\tilde{b} \tilde{c^{\prime}}+\tilde{b} \tilde{d^{\prime}} +\tilde{c^{\second}}\tilde{d^{\prime}}+\tilde{a^{\second}}\tilde{c^{\prime}}+\tilde{b^{\second}}\tilde{c^{\second}} -
(\tilde{a^{\second}} \tilde{c}+\tilde{b^{\second}} \tilde{c} +\tilde{a^{\second}}\tilde{b^{\prime}}+\tilde{b^{\prime}}\tilde{c^{\prime}}+\tilde{b^{\second}}\tilde{d^{\prime}})=\\=&\tilde{b^{\prime}} \tilde{d^{\prime}} +\tilde{c^{\second}}\tilde{d^{\prime}}-
(\tilde{a^{\second}} \tilde{c^{\second}} +\tilde{a^{\second}}\tilde{b^{\prime}}).
\end{align*}
The last term is not necessarily zero. But if one uses the fact that the bicharacters are even, then one has $\tilde{a^{\second}}=\tilde{d^{\prime}}$ and thererore 
$\tilde{b^{\prime}} \tilde{d^{\prime}}=\tilde{a^{\second}}\tilde{b^{\prime}}$ and $\tilde{c^{\second}}\tilde{d^{\prime}}=\tilde{a^{\second}}\tilde{c^{\second}}$. Thus the equlaity \eqref{eq:310}=\eqref{eq:311} holds for nontrivial even bicharacters, which allows for a consistent definition.
\end{remark}
\begin{remark}
The notion of super bicharacter  is similar  to the notion of twist induced by Laplace pairing (or the more general concept of a Drinfeld twist) as  described in \cite{Oeckl}.
\end{remark}
\begin{defn}\begin{bf}(Convolution product)\end{bf}
Let $r$ and $s$ are two even bicharacters on $V$. Define a convolution
product $r\star s$ by
\begin{displaymath}
(r\star s)_{z_1,z_2} (a\ten b)=\sum \ (-1)^{\tilde{a^{\second}} \tilde{b^{\prime}} } r_{z_1,z_2}(a^{\prime}\ten
b^{\prime })s_{z_1,z_2}(a^{\second}\ten b^{\second}).  
\end{displaymath}
The identity bicharacter is given by $r(a\ten b)= \eta (a) \ten \eta
(b) $. 
The  inverse bicharacter \ $r^{-1}$ is defined by 
\begin{displaymath}
r^{-1}_{z_1,z_2}(a\ten b)=r_{z_1,z_2}(S(a)\ten b).
\end{displaymath}
\end{defn}
\begin{lem}
The even bicharacters on $V$ form a supercommutative group with respect to the convolution product with identity and inverse bicharacters given above. 
\end{lem}
\begin{proof}
We will only prove associativity, the rest is proved similarly. Let $a, b\in V$, and $r, s, u$ are even bicharacters on $V$. We have 
\begin{align*}
&((r\star s)\star u)_{z_1,z_2} (a\ten b)=\sum \ (-1)^{\tilde{a^{\second}} \tilde{b^{\prime}} } (r\star s)_{z_1,z_2}(a^{\prime}\ten b^{\prime })u_{z_1,z_2}(a^{\second}\ten b^{\second})=\\ &=\sum \ (-1)^{\tilde{a^{\second}} \tilde{b^{\prime}}+\tilde{(a^{\prime})^{\second}} \tilde{(b^{\prime})^{\prime}} }r_{z_1,z_2}((a^{\prime})^{\prime}\ten (b^{\prime})^{\prime })s_{z_1,z_2}((a^{\prime})^{\second}\ten (b^{\prime})^{\second })u_{z_1,z_2}(a^{\second}\ten b^{\second})=\\ &=\sum \ (-1)^{\tilde{a^{(3)}} \tilde{b^{(1)}}+\tilde{a^{(3)}} \tilde{b^{(2)}}+\tilde{a^{(2)}} \tilde{b^{(1)}} }r_{z_1,z_2}(a^{(1)}\ten b^{(1)})s_{z_1,z_2}(a^{(2)}\ten b^{(2)})u_{z_1,z_2}(a^{(3)}\ten b^{(3)}).
\end{align*}
Here we denote the coassociativity relation $a^{(3)}\ten a^{(3)}\ten a^{(3)}=(1\otimes \Delta)\Delta (a)=(\Delta\otimes 1)\Delta (a)$, for any $a\in V$.
On the other hand, 
\begin{align*}
&(r\star (s\star u))_{z_1,z_2} (a\ten b)=\sum \ (-1)^{\tilde{a^{\second}} \tilde{b^{\prime}} } r_{z_1,z_2}(a^{\prime}\ten b^{\prime })(s\star u)_{z_1,z_2}(a^{\second}\ten b^{\second})=\\ &=\sum \ (-1)^{\tilde{a^{\second}} \tilde{b^{\prime}}+\tilde{(a^{\second})^{\second}} \tilde{(b^{\second})^{\prime}} }r_{z_1,z_2}(a^{\prime}\ten b^{\prime})s_{z_1,z_2}((a^{\second})^{\prime}\ten (b^{\second})^{\prime })u_{z_1,z_2}((a^{\second})^{\second}\ten (b^{\second})^{\second})=\\ &=\sum \ (-1)^{\tilde{a^{(2)}} \tilde{b^{(1)}}+\tilde{a^{(3)}} \tilde{b^{(1)}}+\tilde{a^{(3)}} \tilde{b^{(2)}} }r_{z_1,z_2}(a^{(1)}\ten b^{(1)})s_{z_1,z_2}(a^{(2)}\ten b^{(2)})u_{z_1,z_2}(a^{(3)}\ten b^{(3)}).
\end{align*}
\end{proof}
\begin{defn}\begin{bf}(Transpose and braiding bicharacters)\end{bf}
The transpose of a bicharacter is defined by
\begin{equation}
r^\tau_{z_1,z_2}(a\otimes b)=r_{z_2,z_1}\circ \tau (a\ten b).
\end{equation}
Define a braiding bicharacter $R_{z_1,z_2}$ associated
to $r_{z_1,z_2}$ by
\begin{equation}
  \label{eq:defR}
R_{z_1,z_2}=r\inv_{z_1,z_2}\ast r^\tau_{z_1,z_2}.
\end{equation}
\end{defn}
$R_{z_1,z_2}$ is the obstruction to $r$ being \emph{symmetric}:
$r=r^\tau$.

From now on assume that $V$ is also an $H_D$-module algebra. 
\begin{defn}$(\mathbf{H_D\otimes H_D}$\textbf{-covariant bicharacter)}
In case the bicharacter additionally satisfies :
 \begin{equation} r_{z_1,z_2}(D^ka\otimes D^\ell
  b)=\partial_{z_1}^k\partial_{z_2}^\ell r_{z_1,z_2}(a\otimes b),
\end{equation} for all   $a,b\in V$,
we call the bicharacter \emph{$H_D\otimes H_D$-covariant}.
\end{defn}
\begin{defn}\begin{bf}(Shift bicharacter)\end{bf}
Define for a bicharacter $r_{z_1,z_2}$ a shift
\begin{equation}
  \label{eq:gammabichar}
  r^\gamma_{z_1,z_2}=r_{z_1+\gamma,z_2+\gamma}.
\end{equation}
\end{defn}
The shift $r^\gamma_{z_1,z_2}$ is again a bicharacter. If
$r_{z_1,z_2}$ is $H_D\otimes H_D$-covariant we have the following
expansion:
\[
i_{z_1,z_2;\gamma} r^\gamma_{z_1,z_2}=r_{z_1,z_2}\circ \Delta(e^{\gamma D}).
\]
We can  relate the shift $r^\gamma$
to $r$ by
\begin{equation}
  \label{eq:DefRgamma}
r^\gamma_{z_1,z_2}=r_{z_1,z_2}\ast R^\gamma_{z_1,z_2}, \quad
R^\gamma_{z_1,z_2}=r\inv_{z_1,z_2}\ast r^\gamma_{z_1,z_2}.
\end{equation}
We call $R^\gamma_{z_1,z_2}$ the \emph{translation bicharacter}
associated to $r_{z_1,z_2}$. It is the obstruction to $r$ being shift
invariant (i.e., to $r$ being a function just of $z_1-z_2$).
\begin{notation}
Let $W_2$ be the algebra of power series in $t$, with coefficients
rational functions in $z_1,z_2$ with poles at $z_1=0$ or
$z_1=z_2$ (but \emph{not} at $z_2=0$):
\begin{equation}
  \label{eq:DefW2}
W_2=\mathbb C[z_1^{\pm 1},z_2, (z_1-z_2)^{\pm 1}][[t]].
\end{equation}
\end{notation}
\begin{defn} \label{defn:DefSingmult}\begin{bf}(Singular multiplication map)\end{bf}
Let $V$ be a  Hopf supercommutative supecocommutative superalgebra with antipode $S$, which is also an $H_D$-module algebra. Let $r_{z_1,z_2}$ be an  $H_D\otimes H_D$-covariant bicharacter on $V$ with target space  $W_2$. Define  the  singular multiplication map
\[
X_{z_1,z_2}\colon V^{\otimes 2}\to
V\otimes[[z_1,z_2]][z_1\inv,(z_1-z_2)\inv][[t]], 
\]
by
\begin{equation}
\label{eq:DefSingmult1}
X_{z_1,z_2}(a\otimes b)=\sum (-1)^{\tilde{a^{\prime\prime}} \tilde{b^{\prime}} } (e^{z_1D}a^\prime )(e^{z_2D}b^\prime
)r_{z_1,z_2}(a^{\prime\prime}\otimes b^{\prime\prime}),
\end{equation}
where  $a, b$ are homogeneous elements of $V$. The map $X_{z_1,z_2}$ is extended by linearity to the whole of $V$.
\end{defn}
\begin{defn} \begin{bf}(Braiding and translation maps)\end{bf} We define for any bicharacter $r_{z_1,z_2}$ on $V$ (with target space  $W_2$) a map
$S^{r_{z_1,z_2}}$ on $V\otimes V$ by
\begin{equation}
  \label{eq:defS(rho)}
  S^{r_{z_1,z_2}}(a\otimes b)= \sum (-1)^{\tilde{a^{\prime\prime}} \tilde{b^{\prime}} }a^\prime\otimes b^\prime
  r_{z_1,z_2}(a^{\prime\prime}\otimes b^{\prime\prime}), 
\end{equation}
where  $a, b$ are homogeneous elements of $V$. The map $S^{r_{z_1,z_2}}$ is extended by linearity to the whole of $V$.
In particular, with the braiding bicharacter $R_{z_1,z_2}$ (\eqref{eq:defR}) we associate the map
\begin{equation}
  \label{eq:defStau}
  S^{(\tau)}_{z_1,z_2}=S^{R_{z_1,z_2}}\colon V\otimes V\to   V\otimes
  V[z_1^{\pm1},z_2^{\pm1},(z_1-z_2)^{\pm1}][[t]], 
\end{equation}
and associated to the translation bicharacter \eqref{eq:DefRgamma} we
get a map
\begin{equation}
  \label{eq:defSgama}
  S^{(\gamma)}_{z_1,z_2}=S^{R^\gamma_{z_1,z_2}}\colon V\otimes V\to   V\otimes
  V[z_1^{\pm1},z_2,(z_1+\gamma)^{\pm1}, (z_2+\gamma), (z_1-z_2)^{\pm1}][[t]], 
\end{equation}
\end{defn}
 \begin{thm}\begin{bf}(Super-bicharacter construction)\end{bf}\label{thm:h_d-quantum-vertex-from-bichar}
     Let $V$ be a   supecocommutative Hopf superalgebra with antipode $S$, which is also an $H_D$-module algebra,  and a $H_D\otimes H_D$-covariant super-bicharacter $r_{z_1,z_2}$ with target space  $W_2$. Then the singular multiplication
    $X_{z_1,z_2}$  and the maps $S^{(\tau)}_{z_1,z_2}\circ \tilde{I}$,
    $S^{(\gamma)}_{z_1,z_2}$ defined by (\ref{eq:DefSingmult1}),
    (\ref{eq:defStau}) and (\ref{eq:defSgama}) give $V$ the structure
    of an $H_D$-quantum super vertex algebra as in Definition
    \ref{defn:h_d-quantum-vertex-alg}.
  \end{thm}
\begin{proof}
We will not give the proofs for  all the axioms, as they are similar to \cite{AB}, with  addition of   arguments similar to the argument in remark \ref{remark:even}. We will illustrate it for one of the axioms,  axiom  \eqref{eq:H_DcovMult}: 
\begin{align*}
&e^{\gamma D}X_{z_1,z_2}S^{(\gamma)}_{z_1,z_2}(a\ten b) =e^{\gamma D}X_{z_1,z_2}(\sum (-1)^{\tilde{a^{\prime\prime}} \tilde{b^{\prime}} }a^\prime\otimes b^\prime R^\gamma_{z_1,z_2}(a^{\prime\prime}\otimes b^{\prime\prime}))=\\
&=\sum (-1)^{\tilde{a^{\second}} \tilde{b^{\prime}}+\tilde{(a^{\prime})^{\second}} \tilde{(b^{\prime})^\prime} }(e^{(z_1+\gamma )D}(a^\prime )^\prime )(e^{(z_2+\gamma )D}(b^\prime )^\prime ) r_{z_1, z_2}((a^\prime )^{\second} \ten (b^\prime )^{\second} )R^\gamma_{z_1,z_2}(a^{\prime\prime}\otimes b^{\prime\prime})=\\ &=\sum (-1)^{\tilde{a^{(3)}}(\tilde{b^{(1)}}+\tilde{b^{(2)}}) +\Tilde{a^{(2)}} \tilde{b^{(1)} } }(e^{(z_1+\gamma )D}(a^{(1)})(e^{(z_2+\gamma )D}(b^{(1)}) 
)r_{z_1, z_2}(a^{(2)} \ten b^{(2)} )R^\gamma_{z_1,z_2}(a^{(3)}\otimes b^{(3)})=\\ &=\sum (-1)^{\tilde{(a^{\second})^{\second}}(\tilde{b^{\prime}}+\tilde{(b^{\second})^{\prime}}) +\Tilde{(a^{\second})^{\prime}} \tilde{b^{\prime} } }(e^{(z_1+\gamma )D}(a^{\prime})(e^{(z_2+\gamma )D}(b^{\prime}) 
)r_{z_1, z_2}((a^{\second})^{\prime} \ten (b^{\second})^{\prime} )R^\gamma_{z_1,z_2}((a^{\second})^{\second}\otimes (b^{\second})^{\second})=\\ &=\sum (-1)^{\tilde{(a^{\second})^{\second}}(\tilde{b^{\prime}}+\tilde{(b^{\second})^{\prime}}) +\Tilde{(a^{\second})^{\prime}} \tilde{b^{\prime} } +\Tilde{(a^{\second})^{\second}} \tilde{(b^{\second})^{\prime} } }(e^{(z_1+\gamma )D}(a^{\prime})(e^{(z_2+\gamma )D}(b^{\prime}) 
)(r\star R^\gamma)_{z_1, z_2}(a^{\second} \ten b^{\second} )=\\ &=\sum (-1)^{\tilde{(a^{\second})^{\second}}(\tilde{b^{\prime}}+\tilde{(b^{\second})^{\prime}}) +\Tilde{(a^{\second})^{\prime}} \tilde{b^{\prime} } +\Tilde{(a^{\second})^{\second}} \tilde{(b^{\second})^{\prime} } }(e^{(z_1+\gamma )D}(a^{\prime})(e^{(z_2+\gamma )D}(b^{\prime}) 
)(r\star r^{-1}\star r^{\gamma})_{z_1, z_2}(a^{\second} \ten b^{\second} )=\\ & = X_{z_1+\gamma,z_2+\gamma}
\end{align*}
The proofs that the bicharacter construction satisfies the rest of the axioms are similar, and involve tedious checking of the exponents of $(-1)$ (axioms \eqref{eq:CompatXx1} and \eqref{eq:Compat1xX} are especially unpleasant).
\end{proof}
\begin{cor}
[\textbf{Braided Symmetry}]\label{lem:X2braiding}  For any $a, b\in V$, the singular multiplication
    $X_{z_1,z_2}$  and the map $S^{(\tau)}_{z_1,z_2}$ defined correpondingly by (\ref{defn:DefSingmult}) and 
    (\ref{eq:defStau}) satisfy the braided symmetry relation 
\[
  X_{z_1,z_2}=X_{z_2,z_1}S_{z_2,z_1}^{(\tau)}\circ \tau.
\]
\end{cor}
\begin{proof}
This property generalizes the  important fact of the theory of classical vertex algebras, namely commutativity of  the singular multiplication maps $X_{z_1,z_2}$. We will give a proof of this, as it is similar, but shorter than the Locality axiom, and illustrates why the map $\tilde{I}$ is required in the definition of the braiding map of the $H_D$ quantum vertex algebra as $S^{(\tau)}_{z_1,z_2}\circ \tilde{I}$.
\begin{align*}
&X_{z_2,z_1}S_{z_2,z_1}^{(\tau)}\circ \tau =X_{z_1,z_2}(\sum (-1)^{\tilde{b^{\prime\prime}} \tilde{a^{\prime}}+\tilde{b} \tilde{a} }b^\prime\otimes a^\prime R^\tau_{z_2,z_1}(b^{\prime\prime}\otimes a^{\prime\prime}))=\\ &=\sum (-1)^{\tilde{b^{\prime\prime}} \tilde{a^{\prime}}+\tilde{b} \tilde{a}+\tilde{(b^\prime)^{\second}}\tilde{(a^\prime)^\prime} }(e^{z_2D}(b^\prime)^\prime)(e^{z_1D}(a^\prime)^\prime )r_{z_2, z_1}((b^\prime)^{\second}\otimes (a^\prime)^{\second} ) R^\tau_{z_2,z_1}(b^{\prime\prime}\otimes a^{\prime\prime}))=\\ &=\sum (-1)^{\tilde{b^{\prime\prime}} \tilde{a^{\prime}}+\tilde{b} \tilde{a}+\tilde{(b^\prime)^{\second}}\tilde{(a^\prime)^\prime}+\tilde{(b^{\prime\prime})^{\second}}\tilde{(a^{\prime\prime})^{\prime}} }.\\ &\hspace{0.5cm} .(e^{z_2D}(b^\prime)^\prime)(e^{z_1D}(a^\prime)^\prime )r_{z_2, z_1}((b^\prime)^{\second}\otimes (a^\prime)^{\second} ) (r^{-1}_{z_2,z_1}((b^{\prime\prime})^{\prime}\otimes (a^{\prime\prime})^{\prime}) r^\tau_{z_2,z_1}((b^{\prime\prime})^{\second}\otimes (a^{\prime\prime})^{\second}))=
\end{align*} 
\begin{align*}&=\sum (-1)^{(\tilde{b^{(3)}}+\tilde{b^{(4)}})( \tilde{a^{(1)}}+\tilde{a^{(2)}})+\tilde{b} \tilde{a}+\tilde{b^{(2)}}\tilde{a^{(1)}}+\tilde{b^{(4)}}\tilde{a^{(3)}} }.\\ &\hspace{2cm} .(e^{z_2D}b^{(1)})(e^{z_1D}a^{(1)})  r_{z_2, z_1}(b^{(2)}\otimes a^{(2)} ) r^{-1}_{z_2,z_1}(b^{(3)}\otimes a^{(3)}) r^\tau_{z_2,z_1}(b^{(4)}\otimes a^{(4)})=\\ &=\sum (-1)^{(\tilde{b^{(3)}}+\tilde{b^{(4)}})( \tilde{a^{(1)}}+\tilde{a^{(2)}})+\tilde{b} \tilde{a}+\tilde{b^{(2)}}\tilde{a^{(1)}}+\tilde{b^{(4)}}\tilde{a^{(3)}}+\tilde{b^{(3)}}\tilde{a^{(2)}}+\tilde{b^{(4)}}\tilde{a^{(4)}}+\tilde{b^{(1)}}\tilde{a^{(1)}} }.\\ &\hspace{2cm} .(e^{z_1D}a^{(1)})(e^{z_2D}b^{(1)})  r_{z_1,z_2}(a^{(2)}\otimes b^{(2)})=\\ &=\sum (-1)^{\tilde{a^{(2)}}\tilde{b^{(1)}}}(e^{z_1D}a^{(1)})(e^{z_2D}b^{(1)})  r_{z_1,z_2}(a^{(2)}\otimes b^{(2)})= X_{z_1,z_2}(a\ten b).
\end{align*}
Here we have used the coassociativity of the coproduct and the commutativity of $V$.
\end{proof}
\begin{cor}
\label{cor:Cor}
If the bicharacter $r_{z_1,z_2}(a\ten b)$ as a function of $z_1$ and $z_2$ depends only on $(z_1-z_2)$ for any $a, b\in V$, then the vector superspace  $V$ together with the state-field correspondence given by $Y(a, z)b=X_{z,0}(a\ten b)$ satisfies the axioms for a quantum vertex algebra in the sense of Etingof and Kazhdan,  \cite{EK}.
If further the bicharacter $r_{z_1,z_2}(a\ten b)$  is symmetric, $r=r^\tau$, for any $a, b\in V$, then the vector superspace $V$ is a (nonquantized) super vertex algebra.
\end{cor}
One of the invaluable benefits of the bicharacter construction is the fact that it provides explicit formulas for the braiding and translation maps. (Without the bicharacters, such formulas can only be  given in general  for proper subspaces of the quantum vertex algebras, for example in \cite{FR}. Another solution of this problem is for the braiding map  to be assumed rational, which is in fact what the definition of a quantum vertex algebra in \cite{EK} assumes.)
Even for nonquantized vertex algebras the bicharacter construction has another benefit--there are explicit formulas for the operator product expansions of fields, as well as for the normal ordered products,  in terms of the algebra product on $V$. We also have a formula for  the analytic continuation of a product of arbitrary number of fields. We will start by describing the last of these formulas.
   
We know that for an $H_D$-quantum vertex algebra the  arbitrary products of the vertex operators can be analytically continued, \cite{AB}:
\begin{thm}[\textbf{Analytic Continuation}]\label{thm:analcontn} Let $V$ be an $H_D$-quantum vertex
  algebra. There exists for all $n\ge 2$ maps
\[
X_{z_1,\dots,z_n}\colon V^{\ten n}\to
V[[z_k]][z_j\inv,(z_i-z_j)\inv][[t]],\quad {1\le i<j \le n}, i\le k\le n
\]
such that
\begin{equation}
i_{z_1;z_2,\dots,z_n}X_{z_1,\dots,z_n}=X_{z_1,0}(1\ten
X_{z_2,\dots,z_n}),\label{eq:X1tenXexpansionXn}
\end{equation} 
or in terms of the vertex operators $Y$ we have for any $a_1, a_2, \dots , a_n\in V$ 
\begin{equation}
  \label{eq:FullExpansX}
i_{z_1;z_2;\dots;z_n}X_{z_1,z_2,\dots,z_n}(a_1\ten  a_2\ten  \dots \ten a_n)=Y(a_1,z_1)Y(a_2,z_2)\dots Y(a_n,z_n)1.
\end{equation}
\end{thm}
One of the benefits of the bicharacter construction is that there are explicit formulas for the maps $X_{z_1,\dots,z_n}$ in terms of the bicharacter. We will only give here the formula for $X_{z_1, z_2, z_3}$. 
\begin{lem}[\textbf{Bicharacter formula for the analytic continuation}]
\label{lem:analcont2}
Let $V$ be an $H_D$-quantum vertex algebra defined via a bicharacter $r_{z_1, z_2}$ as in theorem \ref{thm:h_d-quantum-vertex-from-bichar}. We have for any $a, b, c$ homogeneous elements of $V$
\begin{align}
\label{eq:XinBich}
&X_{z_1, z_2, z_3}(a\ten b\ten c)=\sum (-1)^{\tilde{b^{(3)}}(\tilde{c^{(1)}}+\tilde{c^{(2)}}) + (\tilde{a^{(2)}}+ \tilde{a^{(3)}})(\tilde{b^{(1)}}+\tilde{c^{(1)}})+\tilde{a^{(3)}}\tilde{b^{(2)}}+\tilde{b^{(2)}}\tilde{c^{(1)}}}. \nonumber \\ & \hspace{4cm}.(e^{z_1D}a^{(1)})(e^{z_2D}b^{(1)} )(e^{z_1D}c^{(1)} )r_{z_1,z_2}(a^{(2)}\otimes b^{(2)})r_{z_1,z_3}(a^{(3)}\otimes c^{(2)})r_{z_2,z_3}(b^{(3)}\otimes c^{(3)}).
\end{align}
Here as usual we denote $\Delta ^2(a)=a^{(1)}\ten a^{(2)}\ten a^{(3)}$ for any $a\in V$. The map $X_{z_1, z_2, z_3}$ is  extended  to  the whole of $V$ by linearity.
\end{lem}
\begin{proof}
It is enough to prove that 
\begin{align*}
& Y(a,z)Y(b,w)c=i_{z,w}X_{z,w,0}(a\ten b\ten c)=\sum (-1)^{\tilde{b^{(3)}}(\tilde{c^{(1)}}+\tilde{c^{(2)}}) + (\tilde{a^{(2)}}+ \tilde{a^{(3)}})(\tilde{b^{(1)}}+\tilde{c^{(1)}})+\tilde{a^{(3)}}\tilde{b^{(2)}}+\tilde{b^{(2)}}\tilde{c^{(1)}}}.\\ & \hspace{4cm}. (e^{zD}a^{(1)})(e^{wD}b^{(1)} )(c^{(1)} )i_{z,w}r_{z, w}(a^{(2)}\otimes b^{(2)})r_{z, 0}(a^{(3)}\otimes c^{(2)})r_{w, 0}(b^{(3)}\otimes c^{(3)}),
\end{align*}
Note that by assumption for an  $H_D\otimes H_D$-covariant bicharacter we have
\mbox{$i_{z,w}r_{z,w}=(e^{w\partial _x}r_{z,x})\arrowvert _{x=0}$}. Therefore
\begin{align*}
&Y(a,z)Y(b,w)c=\sum Y(a,z)(-1)^{\tilde{b^{\second}} \tilde{c^{\prime}}}((e^{wD}b^{\prime})c^{\prime})r_{w,0}(a^{\second}\ten b^{\second})=\\ &=\sum (-1)^{\tilde{b^{\prime\prime}} \tilde{c^{\prime}}+\tilde{a^{\prime\prime}} \tilde{(b^{\prime}c^{\prime})^{\prime}}}(e^{zD}a^{\prime})((e^{wD}b^{\prime})c^{\prime})^{\prime}r_{z,0}(a^{\second}\ten (e^{wD}(b^{\prime}))c^{\prime})^{\second})r_{w,0}(b^{\second}\ten b^{\second})=
\end{align*}
\begin{align*}
&=\sum (-1)^{\tilde{b^{\prime\prime}} \tilde{c^{\prime}}+\tilde{a^{\prime\prime}} \tilde{(b^{\prime})^{\prime}}+\tilde{a^{\prime\prime}} \tilde{(c^{\prime})^{\prime}}+\tilde{(b^{\prime})^{\second}}\tilde{(c^{\prime})^{\prime}} }(e^{zD}a^{\prime})(e^{wD}(b^{\prime})^{\prime})(c^{\prime})^{\prime}r_{z,0}(a^{\second}\ten (e^{wD}(b^{\prime}))c^{\prime})^{\second})r_{w,0}(b^{\second}\ten b^{\second})=\\
 &=\sum (-1)^{\tilde{(b^{\prime\prime})^{\second}} (\tilde{c^{\prime}}+\tilde{(c^{\second})^{\prime}})+\tilde{a^{\prime\prime}} \tilde{b^{\prime}}+\tilde{a^{\prime\prime}} \tilde{c^{\prime}}+ \tilde{(b^{\second})^{\prime}}\tilde{c^{\prime}}}(e^{zD}a^{\prime})(e^{wD}b^{\prime})(c^{\prime}).\\ & \hspace{4cm} .r_{z,0}(a^{\second}\ten (e^{wD}((b^{\second})^{\prime})(c^{\second})^{\prime}))r_{w,0}((b^{\second})^{\second}\ten (c^{\second})^{\second})=\\ &=\sum (-1)^{\tilde{(b^{\prime\prime})^{\second}} (\tilde{c^{\prime}}+\tilde{(c^{\second})^{\prime}})+(\tilde{(a^{\prime\prime})^{\prime}}+\tilde{(a^{\prime\prime})^{\second}}) (\tilde{b^{\prime}}+\tilde{c^{\prime}})+ \tilde{(b^{\second})^{\prime}}\tilde{c^{\prime}}+\tilde{(a^{\prime\prime})^{\second}} \tilde{(b^{\second})^{\prime}}}(e^{zD}a^{\prime})(e^{wD}b^{\prime})(c^{\prime}).\\ & \hspace{4cm} .r_{z,0}((a^{\second})^{\prime}\ten (e^{wD}(b^{\second})^{\prime}))r_{z,0}((a^{\second})^{\second}\ten (c^{\second})^{\prime}))r_{w,0}((b^{\second})^{\second}\ten (c^{\second})^{\second})=\\
 &=\sum (-1)^{\tilde{(b^{\prime\prime})^{\second}} (\tilde{c^{\prime}}+\tilde{(c^{\second})^{\prime}})+(\tilde{(a^{\prime\prime})^{\prime}}+\tilde{(a^{\prime\prime})^{\second}}) (\tilde{b^{\prime}}+\tilde{c^{\prime}})+ \tilde{(b^{\second})^{\prime}}\tilde{c^{\prime}}+\tilde{(a^{\prime\prime})^{\second}} \tilde{(b^{\second})^{\prime}}}(e^{zD}a^{\prime})(e^{wD}b^{\prime})(c^{\prime}).\\ & \hspace{4cm} .(e^{w\partial _x}r_{z,x}((a^{\second})^{\prime}\ten (b^{\second})^{\prime}))\arrowvert _{x=0}r_{z,0}((a^{\second})^{\second}\ten (c^{\second})^{\prime}))r_{w,0}((b^{\second})^{\second}\ten (c^{\second})^{\second})=\\ &=\sum (-1)^{\tilde{(b^{\prime\prime})^{\second}} (\tilde{c^{\prime}}+\tilde{(c^{\second})^{\prime}})+(\tilde{(a^{\prime\prime})^{\prime}}+\tilde{(a^{\prime\prime})^{\second}}) (\tilde{b^{\prime}}+\tilde{c^{\prime}})+\tilde{(b^{\second})^{\prime}}\tilde{c^{\prime}}+\tilde{(a^{\prime\prime})^{\second}} \tilde{(b^{\second})^{\prime}}}(e^{zD}a^{\prime})(e^{wD}b^{\prime})(c^{\prime}).\\ & \hspace{4cm} .(i_{z,w}r_{z,w}((a^{\second})^{\prime}\ten (b^{\second})^{\prime}))r_{z,0}((a^{\second})^{\second}\ten (c^{\second})^{\prime}))r_{w,0}((b^{\second})^{\second}\ten (c^{\second})^{\second})=\\ &=i_{z,w}X_{z,w,0}(a\ten b\ten c).
\end{align*}
\end{proof}
Similar formulas can be derived for any $X_{z_1,\dots,z_n}$, $n\in \mathbf{N}$.
The advantage of the formula in lemma \ref{lem:analcont2} is that even though it seems long, it is imminently amenable to Laurent  expansions, as the singularity in $z, w$ depends only on the bicharacter, which for any $a, b\in V$ is just an ordinary function of $z, w$.
\begin{thm}\begin{bf}(Bicharacter formula for the residues)\end{bf}
\label{thm:comlH_{D, T}}
Assume that $V$ is endowed with a quantum vertex algebra structure given by a  bicharacter $r_{z, w}$. Further,  assume that for any $a, b\in V$ the bicharacter $r_{z,w}(a\ten b)$ is a meromorphic function and can be expanded around $z=w$ as \mbox{$r_{z,w}(a\ten b)=\sum_{k=0}^{N-1}\frac{f^k_{a,b}}{(z- w)^{k+1}} +reg.$}, $N=N_{a, b}$ is  the order of the pole at $z=w$. 
For any $a,\ b \in V$ and any $n\in \mathbf{N}, n\le N$ we have 
\begin{equation}
\Res_{z= w}X_{z,w,0}(a\ten b\ten c)(z- w)^n dz=\sum_{k=n}^{N-1}\sum (-1)^{\tilde{a^{\second}}\tilde{b^{\prime}}}f^k_{a^{\second}, b^{\second}} Y((D^{(k-n)}a^{\prime }).b^{\prime }, w)c.
\end{equation}
\end{thm}
\begin{proof}
By using coassociativity and cocommutativity we have from \eqref{eq:XinBich}
\begin{align*}
&X_{z,w,0}(a\ten b\ten c)=\sum (-1)^{\tilde{(b^{\second})^{\second}}(\tilde{c^{\prime}}+\tilde{(c^{\second})^{\prime}}) + (\tilde{(a^{\second})^{\prime}}+ \tilde{(a^{\second})^{\second}})(\tilde{b^{\prime}}+\tilde{c^{\prime}})+\tilde{(a^{\second})^{\second}}\tilde{(b^{\second})^{\prime}}+\tilde{(b^{\second})^{\prime}}\tilde{c^{\prime}} }.\\ & \hspace{4cm} .(e^{zD}a^{\prime})(e^{wD}b^{\prime})c^{\prime}r_{z,w}((a^{\second})^{\prime}\ten (b^{\second})^{\prime}))r_{z,0}((a^{\second})^{\second}\ten c^{\second})^{\prime})r_{w,0}((b^{\second})^{\second}\ten (c^{\second})^{\second})=\\
 &=\sum (-1)^{\tilde{(b^{\second})^{\prime}}(\tilde{c^{\prime}}+\tilde{(c^{\second})^{\prime}}) + (\tilde{(a^{\second})^{\second}}+ \tilde{(a^{\second})^{\prime}})(\tilde{b^{\prime}}+\tilde{c^{\prime}})+\tilde{(a^{\second})^{\prime}}\tilde{(b^{\second})^{\second}}+\tilde{(b^{\second})^{\second}}\tilde{c^{\prime}}+\tilde{(a^{\second})^{\prime}}\tilde{(a^{\second})^{\second}}+\tilde{(b^{\second})^{\prime}}\tilde{(b^{\second})^{\second}} }(e^{zD}a^{\prime})(e^{wD}b^{\prime})c^{\prime}.\\ & \hspace{4cm} .r_{z,w}((a^{\second})^{\second}\ten (b^{\second})^{\second}))r_{z,0}((a^{\second})^{\prime}\ten c^{\second})^{\prime})r_{w,0}((b^{\second})^{\prime}\ten (c^{\second})^{\second})=\\
 &=\sum (-1)^{\tilde{(b^{\prime})^{\second}}(\tilde{c^{\prime}}+\tilde{(c^{\second})^{\prime}}) + (\tilde{a^{\second}}+ \tilde{(a^{\prime})^{\second}})(\tilde{(b^{\prime})^{\prime}}+\tilde{c^{\prime}})+\tilde{(a^{\prime})^{\second}}\tilde{b^{\second}}+\tilde{b^{\second}}\tilde{c^{\prime}}+\tilde{(a^{\prime})^{\second}}\tilde{a^{\second}}+\tilde{(b^{\prime})^{\second}}\tilde{b^{\second}}}e^{zD}(a^{\prime}))^{\prime }e^{wD}((b^{\prime}))^{\prime }c^{\prime}).\\ & \hspace{4cm} .r_{z,w}(a^{\second}\ten b^{\second}))r_{z,0}((a^{\prime })^{\second }\ten (c^{\second})^{\prime})r_{w,0}((b^{\prime })^{\second}\ten (c^{\second})^{\second}).
\end{align*}
Note that $r_{z,0}((a^{\prime })^{\second }\ten (c^{\second})^{\prime})$ as a function of $z$ is regular at $z= w$, and therefore can be expanded in a power series in $(z- w)$:
\begin{align*}
r_{z,0}((a^{\prime })^{\second }\ten (c^{\second})^{\prime})&=\sum_{i\ge 0} \big( (\partial _z)^{(i)}r_{z,0}((a^{\prime })^{\second }\ten c^{\second})^{\prime}))\big) \arrowvert _{z= w}(z- w)^i=\sum_{i\ge 0}  r_{w,0}(D^{(i)}(a^{\prime })^{\second }\ten c^{\second})^{\prime}))(z- w)^i
\end{align*}
We have used above the fact that the  bicharacter is $H_D\ten H_D$-covariant. 
\begin{align*}
&\Res_{z= w}X_{z,w,0}(a\ten b\ten c)(z- w)^n= \sum (-1)^{\tilde{(b^{\prime})^{\second}}(\tilde{c^{\prime}}+\tilde{(c^{\second})^{\prime}}) + (\tilde{a^{\second}}+ \tilde{(a^{\prime})^{\second}})(\tilde{(b^{\prime})^{\prime}}+\tilde{c^{\prime}})+\tilde{(a^{\prime})^{\second}}\tilde{b^{\second}}+\tilde{b^{\second}}\tilde{c^{\prime}}+\tilde{(a^{\prime})^{\second}}\tilde{a^{\second}}+\tilde{(b^{\prime})^{\second}}\tilde{b^{\second}}}.\\ &.
r_{w,0}((b^{\prime })^{\second}\ten (c^{\second})^{\second})Res_{z= w} \big((e^{zD}(a^{\prime})^{\prime })(e^{wD}(b^{\prime})^{\prime})c^{\prime}(\sum_i  r_{w,0}(D^{(i)}(a^{\prime })^{\second }\ten c^{\second})^{\prime}))(z- w)^{i+n})r_{z,w}(a^{\second}\ten b^{\second}))\big) =\\
 &=\sum (-1)^{\tilde{(b^{\prime})^{\second}}(\tilde{c^{\prime}}+\tilde{(c^{\second})^{\prime}}) + (\tilde{a^{\second}}+ \tilde{(a^{\prime})^{\second}})(\tilde{(b^{\prime})^{\prime}}+\tilde{c^{\prime}})+\tilde{(a^{\prime})^{\second}}\tilde{b^{\second}}+\tilde{b^{\second}}\tilde{c^{\prime}}+\tilde{(a^{\prime})^{\second}}\tilde{a^{\second}}+\tilde{(b^{\prime})^{\second}}\tilde{b^{\second}}}r_{w,0}((b^{\prime })^{\second}\ten (c^{\second})^{\second}).\\ &. Res_{z= w} \big(e^{(z- w)D}(e^{wD}(a^{\prime})^{\prime})(e^{wD}(b^{\prime})^{\prime})c^{\prime}(\sum_i  r_{w,0}(D^{(i)}(a^{\prime })^{\second }\ten c^{\second})^{\prime}))(z- w)^{i+n})r_{z,w}(a^{\second}\ten
 b^{\second}))\big)  =\\
 &=\sum (-1)^{\tilde{(b^{\prime})^{\second}}(\tilde{c^{\prime}}+\tilde{(c^{\second})^{\prime}}) + (\tilde{a^{\second}}+ \tilde{(a^{\prime})^{\second}})(\tilde{(b^{\prime})^{\prime}}+\tilde{c^{\prime}})+\tilde{(a^{\prime})^{\second}}\tilde{b^{\second}}+\tilde{b^{\second}}\tilde{c^{\prime}}+\tilde{(a^{\prime})^{\second}}\tilde{a^{\second}}+\tilde{(b^{\prime})^{\second}}\tilde{b^{\second}}}r_{w,0}((b^{\prime })^{\second}\ten (c^{\second})^{\second}). \\  & .Res_{z= w} \big( \sum _{k=0}^{N-1}\sum _{i,j\ge 0}D^{(j)}(e^{ wD}(a^{\prime})^{\prime })(e^{wD}(b^{\prime})^{\prime})c^{\prime} r_{w,0}(D^{(i)}(a^{\prime})^{\second }\ten c^{\second})^{\prime}))(z- w)^{i+j+n-k-1}f_{a^{\second}, b^{\second}}^k)\big)  =
\end{align*}
\begin{align*}
 &=\sum (-1)^{\tilde{(b^{\prime})^{\second}}(\tilde{c^{\prime}}+\tilde{(c^{\second})^{\prime}}) + (\tilde{a^{\second}}+ \tilde{(a^{\prime})^{\second}})(\tilde{(b^{\prime})^{\prime}}+\tilde{c^{\prime}})+\tilde{(a^{\prime})^{\second}}\tilde{b^{\second}}+\tilde{b^{\second}}\tilde{c^{\prime}}+\tilde{(a^{\prime})^{\second}}\tilde{a^{\second}}+\tilde{(b^{\prime})^{\second}}\tilde{b^{\second}}}.\\ &.\Big(\sum _{k=n}^{N-1}\sum_{i+j=k-n}\sum D^{(j)}(e^{wD}(a^{\prime})^{\prime }) e^{wD}((b^{\prime})^{\prime })c^{\prime})r_{w,0}(D^{(i)}(a^{\prime })^{\second }\ten c^{\second})^{\prime}))r_{w,0}((b^{\prime })^{\second }\ten c^{\second})^{\second})) )\big) f_{a^{\second}, b^{\second}}^k\Big) = \\ &=\sum (-1)^{\tilde{(b^{\prime})^{\second}}(\tilde{c^{\prime}}+\tilde{(c^{\second})^{\prime}}) + (\tilde{a^{\second}}+ \tilde{(a^{\prime})^{\second}})(\tilde{(b^{\prime})^{\prime}}+\tilde{c^{\prime}})+\tilde{(a^{\prime})^{\second}}\tilde{b^{\second}}+\tilde{b^{\second}}\tilde{c^{\prime}}+\tilde{(a^{\prime})^{\second}}\tilde{a^{\second}}+\tilde{(b^{\prime})^{\second}}\tilde{b^{\second}}}.\\ & .\sum _{k=n}^{N-1} \sum_{i+j=k-n}(e^{wD}(D^{(j)}(a^{\prime})^{\prime })) (e^{wD}(b^{\prime})^{\prime })c^{\prime})r_{w,0}(D^{(i)}(a^{\prime })^{\second }\ten c^{\second})^{\prime}))r_{w,0}((b^{\prime })^{\second }\ten c^{\second})^{\second})) )\big) f_{a^{\second }, b^{\second }}^k  =\\
 &=\sum (-1)^{\tilde{(b^{\prime})^{\second}}(\tilde{c^{\prime}}+\tilde{(c^{\second})^{\prime}}) + (\tilde{a^{\second}}+ \tilde{(a^{\prime})^{\second}})(\tilde{(b^{\prime})^{\prime}}+\tilde{c^{\prime}})+\tilde{(a^{\prime})^{\second}}\tilde{b^{\second}}+\tilde{b^{\second}}\tilde{c^{\prime}}+\tilde{(a^{\prime})^{\second}}\tilde{a^{\second}}+\tilde{(b^{\prime})^{\second}}\tilde{b^{\second}}}.\\ &.\sum _{k=n}^{N-1} (e^{wD}((D^{(k-n)}a^{\prime})^{\prime })) (e^{wD}(b^{\prime})^{\prime })c^{\prime})r_{w,0}(D^{(k-n)}a^{\prime })^{\second }\ten c^{\second})^{\prime}))r_{w,0}((b^{\prime })^{\second }\ten c^{\second})^{\second})) )\big) f_{a^{\second }, b^{\second }}^k  =\\ &= \sum (-1)^{\tilde{(b^{\prime})^{\second}}\tilde{c^{\prime}} + \tilde{a^{\second}}(\tilde{(b^{\prime})^{\prime}}+\tilde{c^{\prime}})+ \tilde{(a^{\prime})^{\second}}\tilde{c^{\prime}}+\tilde{(a^{\prime})^{\second}}\tilde{b^{\second}}+\tilde{b^{\second}}\tilde{c^{\prime}}+\tilde{(a^{\prime})^{\second}}\tilde{a^{\second}}+\tilde{(b^{\prime})^{\second}}\tilde{b^{\second}}}.\\ &.\Big( \sum_{k=n}^{N-1} e^{wD}\big( ((D^{(k-n)}a^{\prime})(b^{\prime}))^{\prime }c^{\prime}\big) )r_{w,0}((D^{(k-n)}a^{\prime }b^{\prime })^{\second}\ten c^{\second})f^k_{a^{\second }, b^{\second }}\Big)=\\ &= \sum_{k=n}^{N-1} \sum (-1)^{(\tilde{(b^{\prime})^{\second}}+\tilde{(a^{\prime}})\tilde{c^{\prime}} + \tilde{a^{\second}}\tilde{b^{\prime}}} e^{wD}\big( ((D^{(k-n)}a^{\prime})(b^{\prime}))^{\prime }c^{\prime}\big) )r_{w,0}((D^{(k-n)}a^{\prime }b^{\prime })^{\second}\ten c^{\second})f^k_{a^{\second }, b^{\second }}=\\ & =\sum_{k=n}^{N-1} \sum (-1)^{\tilde{a^{\second}}\tilde{b^{\prime}}}f^k_{a^{\second },b^{\second }}Y((D^{(k-n)}a^{\prime }).b^{\prime }, w)c.
\end{align*}
Here we took into account that $f^k_{a^{\second },b^{\second }}=0$ unless $\tilde{a^{\second }}=\tilde{b^{\second }}$, as the bicharacters are even.
\end{proof}
\begin{cor}\begin{bf}(Bicharacter formula for the operator product expansion)\end{bf}
\label{cor:opes}
For any $a,\ b \in V$ we have 
\begin{equation}
Y(a,z)Y(b,w)=i_{z,w}\sum_{n=0}^{N-1}  \frac{\sum_{k=n}^{N-1}\sum(-1)^{\tilde{a^{\second}}\tilde{b^{\prime}}}f^k_{a^{\second },b^{\second }}Y((D^{(k)}a^{\prime }).b^{\prime }, w)}{(z- w)^{n+1}} \ \ +\text{regular}.
\end{equation}
\end{cor}
\begin{cor}\begin{bf}(Bicharacter formula for the normal ordered products)\end{bf}
\label{cor:normord}
Assume as above that $V$ is endowed with a quantum vertex algebra structure  given by a  bicharacter $r_{z, w}$.  Further,  let $r_{z, w}$ have around $z=w$ the expansions \mbox{$r_{z,w}(a\ten b)=\sum_{k=-1}^{N-1}\frac{f^k_{a,b}}{(z- w)^{k+1}}$}, $N=N_{a, b}$ is  the order of the pole at $z=w$.  For any $a,\ b \in V$  we have 
\begin{equation}
:Y(a,z)Y(b,z):_{z=w}=\sum_{k=-1}^{N-1}\sum(-1)^{\tilde{a^{\second}}\tilde{b^{\prime}}}f^{k}_{a^{\second },b^{\second }}Y((D^{(k+1)}a^{\prime }).b^{\prime }, z), 
\end{equation}
where $:Y(a,z)Y(b,z):_{z=w}$ is the normal ordered product of the fields $Y(a, z)$ and $Y(b, w)$ around $z=w$. 
\end{cor}
\begin{proof} We would like to recall that the normal ordered product of the fields $Y(a, z)$ and $Y(b, w)$ is defined differently for (nonquantized) super vertex algebras in  \cite{Kac}, but it has the property (Theorem 2.3, \cite{Kac}): 
\begin{equation}
:Y(a,z)Y(b,z):= \Big(Y(a,z)Y(b,w)-i_{z,w}\sum_{n=0}^{N-1}  \frac{a(w)_{(j)}b(w)}{(z- w)^{n+1}}\Big)\arrowvert _{z=w},
\end{equation}
i.e., it is the constant term of the regular part of the operator product expansion of $Y(a,z)Y(b,w)$ around the only singularity $z=w$. Now for quantum vertex algebras there could be other singularities, but as a meromorphic function $r_{z, w}$, and thus $Y(a,z)Y(b,w)$,  have Laurent  expansions  around each one of them. The normal ordered product $:Y(a,z)Y(b,z):_{z=w}$ denotes the constant term in the regular part of the expansion around $z=w$. 
The proof then follows the same lines as the proof of theorem \ref{thm:comlH_{D, T}}, but we need the constant terms $f^{-1}_{a,b}$ of $r_{z,w}(a\ten b)$ as well.
\end{proof}
\begin{remark}
We can allow essential singularities in the expansion of the bicharacter  \mbox{$r_{z,w}(a\ten b)$} in powers of  $(z-w)$, i.e.,  as \mbox{$r_{z,w}(a\ten b)=\sum_{k=0}^{\infty}\frac{f^k_{a,b}}{(z- w)^{k+1}} +reg.$}. The difference is that we would have then an infinite sum in the operator product expansion. An example of such a situation would be if for instance $r_{z,w}(a\ten b)=e^{\frac{t}{z-w}}$ for some $a, b\in V$. This is in fact within the definition of a bicharacter as taking values in $W_2=\mathbb C[z_1^{\pm 1},z_2, (z_1-z_2)^{\pm 1}][[t]]$.
\end{remark}
\vspace{0.6cm}

\section{Bicharacter presentation of the charged free fermion super vertex algebra}
It is well known (see e.g. \cite{Kac}) that  the charged free fermion super vertex algebra can be  described as follows.
Let $\mathit{Cl}$ be the Clifford algebra with generators $\{ \phi _m | m\in \mathbf{Z}\}$ and $\{ \psi_n | n\in \mathbf{Z}\}$, and relations 
\begin{equation}
[\psi^+_m,\psi^-_n]_{\dag}=\delta _{m+n,1}1, \quad [\psi^+_m,\psi^+_n]_{\dag}=[\psi^-_m,\psi^-_n]_{\dag}=0.
\end{equation}
Denote by $\mathit{\Lambda}$ the fermionic Fock space representation of $\mathit{Cl}$ generated by a vector $|0\rangle $,  such that 
\begin{equation}
\psi^+ _n|0\rangle=\psi^- _n|0\rangle=0 \quad \text{for} \quad n>0.
\end{equation}  
The charged free fermion super vertex algebra on  $\mathit{\Lambda}$ is generated by the fields $Y(|0\rangle, z)=Id$, \ $\phi (z)=\sum_{n\in \mathbf{Z}}  \psi^+_n z^{-n}$ and $\psi (z)=\sum_{n\in \mathbf{Z}}  \psi^-_n z^{-n}$. The anticommutation relations imply that the only nontrivial operator product expansion between the generating fields is $\phi (z)\psi (w)\sim \frac{1}{z-w}\sim \psi(z)\phi(w)$. (For definitions and notation one can see \cite{Kac}. Here we prefer to use the  notation  $\phi (z)$ instead of $\psi^+ (z)$, etc.,  for convenience in what follows.)

The fermionic Fock space $\mathit{\Lambda}$ can be given a Hopf super algebra structure in the following way. Denote  $\phi _n:=\psi^+ _{-n}|0\rangle$, \ for $n\ge 0$, and $\psi _n:=\psi^- _{-n}|0\rangle$, \ for $n\ge 0$. 
The algebra structure is determined by requiring that $\phi _n$ and $\psi _m$ are odd and anticommute for all $m, n\ge 0$ (i.e., $\mathit{\Lambda}$ is the exterior algebra with generators $\phi _n$ and $\psi _m$). The  Hopf algebra structure is determined by requiring that $\phi _n$ and $\psi _m$ be primitive for any $m, n\ge 0$ (i.e. $\del (\phi _n)= \phi _n\ten 1 +1\ten \phi _n$, \ $\eta (\phi _n)=0$, \ $S(\phi _n)=-\phi _n$  for any $n\ge 0$, same for $\psi _n$). Moreover $\mathit{\Lambda}$ is an $H_D$ module algebra by $D^{(n)}\phi_0=\phi _n$ and  $D^{(n)}\psi_0=\psi _n$.

From Theorem \ref{thm:h_d-quantum-vertex-from-bichar} we know that any even $H_D\otimes H_D$-covariant bicharacter on $\mathit{\Lambda}$ will give  a structure of $H_D$-quantum vertex algebra on $\mathit{\Lambda}$. 
The requirement that the bicharacter be $H_D\otimes H_D$-covariant implies that we need  (and could) only choose the bicharacter on the elements $\phi _0$ and $\psi _0$ of $\mathit{\Lambda}$, the bicharacter on the rest of $\mathit{\Lambda}$ will be determined automatically by the properties of a bicharacter (see \ref{defn:bich}).
In particular the Corollary \ref{cor:Cor} suggests the following Theorem:
\begin{thm} \ \label{thm:chargedfreeferm}
\begin{itemize} \item The   bicharacter on the  Hopf superalgebra $\mathit{\Lambda}$ defined by 
\begin{align}
&r_{z,w}(\phi_0\ten \psi_0)=\frac{1}{z-w},  \quad r_{z,w}(\psi_0\ten \phi_0)=\frac{1}{z-w}, \\
&r_{z,w}(\phi_0\ten \phi_0)=0,  \quad r_{z,w}(\psi_0\ten \psi_0)=0
\end{align}
gives  $\mathit{\Lambda}$  exactly the structure of the charged free fermion super vertex algebra.
\item The  supecocommutative Hopf superalgebra $\mathit{\Lambda}$ with an even bicharacter $r_{z, w}$ which is a function only of $(z-w)$,  is a quantization of the charged free fermion vertex algebra in the sense of Etingof-Kazhdan ( \cite{EK}).
\item The  supecocommutative Hopf superalgebra $\mathit{\Lambda}$ with any even  bicharacter $r_{z, w}$  is a quantization of the charged free fermion vertex algebra in the sense of an $H_D$-quantum vertex algebra.
\end{itemize}
\end{thm}
\begin{proof}
For the first statement of  the theorem, one observes that the bicharacter is symmetric: 
\begin{equation*}
r_{z, w}^{\tau }(\phi_0\ten \psi_0)=-r_{w,z}(\psi_0\ten \phi_0)=-\frac{1}{w-z}=r_{z,w}(\phi_0\ten \psi_0),
\end{equation*}
therefore this bicharacter defines a (nonquantized) super vertex algebra according to theorem \ref{thm:h_d-quantum-vertex-from-bichar}. To verify which super vertex  algebra this  is in particular,  one has to check  the operator product expansions of the generating vertex operators $Y(\phi_0 , z)$ and $Y(\psi_0 , w)$. We do that using corollary \ref{cor:opes}. We have $r_{z,w}(\phi_0\ten \psi_0)=\frac{1}{z-w}$, and of course $r_{z,w}(1\ten 1)=1=\text{regular}$, \ \mbox{$r_{z,w}(\phi_0\ten 1)=r_{z,w}(1\ten \psi_0)=0$}. Thus the only nonzero $f^k_{\phi_0^{\second },\psi_0^{\second }}$ is $f^0_{\phi_0, \psi_0}=1$  That immediately shows $Y(\phi_0 , z)Y(\psi_0 , w)\sim \frac{1}{z-w}$.\\
In this particularly simple case one can calculate the operator product expansion directly, using  lemma \ref{lem:analcont2}. We will show it as an illustration  of calculations using the bicharacter construction.
We have \begin{align*}
\Delta ^2(\phi_0)&=\phi_0 \ten 1 \ten 1 +1\ten  \phi_0 \ten 1 +1\ten 1 \ten \phi_0 \\ \Delta ^2(\psi_0)&=\psi_0 \ten 1 \ten 1 +1\ten  \psi_0 \ten 1 +1\ten 1 \ten \psi_0.
\end{align*}  
Let $c$ be arbitrary homogeneous element of $\mathit{\Lambda}$, and denote \mbox{$\Delta ^2(c)=c^{(1)}\ten c^{(2)}\ten c^{(3)}$}.
\begin{align*}
Y(\phi_0 , z)Y(\psi_0 , w)c& = c.r_{z, w}(\phi _0\ten \psi _0) +(e^{zD}\phi _0)(e^{wD}\psi _0)c+ (-1)^{\tilde{c^{\prime}}}(e^{zD}\phi _0)c^{\prime}. r_{w, 0}(\psi _0\ten c^{\second})  -\\ & -(-1)^{\tilde{c^{\prime}}}(e^{wD}\psi _0)c^{\prime}. r_{z, 0}(\phi _0\ten c^{\second}) +(-1)^{\tilde{c^{(2)}}}c^{(1)}. r_{z, 0}(\phi _0\ten c^{(2)})r_{w, 0}(\psi _0\ten c^{(3)})=\\ &=\frac{c}{z-w}\ \ +\text{regular},
\end{align*}
which is  proves the first statement. The other two statements are then immediate consequences of corollary \ref{cor:Cor}.
One also can observe that the regular part of the above expansion when evaluated at $z=w$ is the normal ordered product $:Y(\phi_0 , w)Y(\psi_0 , w):c$ (which one  can of course also get from the formula in corollary \ref{cor:normord}):
\begin{align*}
:Y(\phi_0 , w)Y(\psi_0 , w):c& = (e^{wD}\phi _0)(e^{wD}\psi _0)c+ (-1)^{\tilde{c^{\prime}}}(e^{wD}\phi _0)c^{\prime}. r_{w, 0}(\psi _0\ten c^{\second})  -\\ & -(-1)^{\tilde{c^{\prime}}}(e^{wD}\psi _0)c^{\prime}. r_{w, 0}(\phi _0\ten c^{\second}) +(-1)^{\tilde{c^{(2)}}}c^{(1)}. r_{w, 0}(\phi _0\ten c^{(2)})r_{w, 0}(\psi _0\ten c^{(3)})=\\ &=(e^{wD}\phi _0\psi _0)c+ (-1)^{\tilde{c^{\prime}}}(e^{wD}\phi _0)c^{\prime}. r_{w, 0}(\psi _0\ten c^{\second}) -\\ &-(-1)^{\tilde{c^{\prime}}}(e^{wD}\psi _0)c^{\prime}. r_{w, 0}(\phi _0\ten c^{\second}) +c^{\prime}. r_{w, 0}(\phi _0\psi _0\ten c^{\second})=\\ &=Y(\phi_0 \psi_0 , w)c.
\end{align*}
Here we have used that 
\begin{equation*}
\Delta (\phi_0 \psi_0)=\phi_0 \psi_0\ten 1 +1\ten \phi_0 \psi_0 +\phi_0 \ten \psi_0 - \psi_0 \ten \phi_0, 
\end{equation*}
and the definition \eqref{defn:DefSingmult}.
It is well known that the field $:Y(\phi_0 , z)Y(\psi_0 , z):=Y(\phi_0 \psi_0 , z)$ is a Heisenberg field, which one can see from it's operator product expansion with itself, by using corollary \ref{cor:opes}. 
\end{proof}

\def\cprime{$'$}
\providecommand{\bysame}{\leavevmode\hbox to3em{\hrulefill}\thinspace}
\providecommand{\MR}{\relax\ifhmode\unskip\space\fi MR }
\providecommand{\MRhref}[2]{%
  \href{http://www.ams.org/mathscinet-getitem?mr=#1}{#2}
}
\providecommand{\href}[2]{#2}

\end{document}